\documentclass[a4paper,12pt]{article}
\usepackage{graphicx}
\usepackage{epstopdf}
\def\be{\begin{equation}}
\def\ee{\end{equation}}

\begin{document}

\title{Evolution of the density parameter in the anisotropic DGP  cosmology}
\author{ Rizwan Ul Haq Ansari  $^{\dagger \dagger }$\footnote{Email: rizwan@iucaa.ernet.in} \ and P  K Suresh  \footnote{Email: pkssp@uohyd.ernet.in} $^{\star }$\\
\small $^{\dagger \dagger}$  Inter University Center for Astronomy and Astrophysics, \\ \small Post Bag - 4, Ganeshkhind, Pune-411 007. India\\
\small $^{\dagger }$School of Physics, University of Hyderabad, Hyderabad 500 046. India. }

\maketitle

\begin{center}
\bf Abstract
\end{center}
{\small Evolution of the density parameter in the anisotropic DGP braneworld model is studied.  The role of shear and cross-over scale in the evolution of $\Omega_\rho$ is examined for both the branches of solution in the DGP model. The evolution is modified significantly compared to the FRW model and further  it  does not depend on the value of $\gamma$ alone.  Behaviour of the cosmological density parameter $\Omega_\rho$ is unaltered  in the late universe.  The study of decceleration parameter shows that the entry of the universe into self accelerating phase  is determined by the value of shear. We also obtain an estimate of the shear parameter  $\frac{\Sigma}{H_0} \sim 1.68 \times 10^{-10}$, which is in agreement with the constraints obtained in the literature using data.}

\section{Introduction}

Results from the recent observations of the type-I supernovae and WMAP suggest that  we are living in more or less flat  type universe \cite{reiss,perl,spergel,komats}.
These implies that the cosmological density parameter $\Omega$  must be equal to unity or very much near to it. The inflationary scenario also supports  the   value of the density parameter which is very much close to unity.
For the estimation of the  cosmological density parameter to  be unity it is  assumed that apart from the dark energy and baryonic matter the universe  also contains  dark matter. Therefore  the contents of the universe are  of paramount relevance in understanding the  geometry and evolution of the universe. Hence  determination of the density parameter  is very important  because it can  put  constraints on cosmological model building and that are subjected to observations. In general the density parameter can be considered as dynamical quantity rather than a constant.  For the evolutionary study of  the density parameter   one has to take into account  the role of  the effective equation of the state of the epoch of the universe  under consideration. Most of the recent observations  estimate the value of total density parameter $\Omega_0$ to be close to unity, assuming  the homogeneity and isotropy of the spacetime. However, one can also consider the universe that deviates from such idealized situation, to accommodate an anisotropic spacetime. Historically, anisotropic models of the universe have been studied to avoid assumptions of specific initial conditions in the Friedmann-Robertson-Walker (FRW)
models and also  it is believed that  dynamics of  early universe  is profoundly
influenced in  presence of  anisotropy just below
the Planck or string scale \cite{blhu,Maca}. Bianchi models are also relevant in the context of Belinskii-Khalatnikov-Lifshitz (BKL) conjecture \cite{bel}
 which states that in the Einstein
equations `terms containing time derivatives will dominate over those containing spatial derivatives' as one approaches a space-like singularity, such a model is described by homogeneous and anisotropic cosmologies.
Evolution of the  cosmological density parameter  in presence of anisotropy is studied and showed that the initial conditions are robust with respect to shear and viscosity \cite{hensk}.  In the present work our aim is to examine  dynamical behaviour of the density parameter in the anisotropic Dvali-Gabadadze-Porrati (DGP) braneworld cosmology, which  has the potential
to  explain the  current accelerating phase of the universe. The accelerated expansion of the universe is again suggested by the recent observations such as type-I supernovae and WMAP\cite{reiss,perl,spergel,komats}. The DGP braneworld cosmological model has also been studied from
observational point of view and these have put constraints on model
parameters \cite{Lomb,sohrab}.

 The DGP  braneworld cosmological model  is basically  inspired by higher dimensional
theories that have been studied in great detail in recent
times. In these models our observed  four dimensional (4D) universe
is a three dimensional hypersurface known as \textit{brane} embedded in a  higher
dimensional spacetime called \textit{bulk} (for a review of brane cosmology see \cite{brax}).
The application of such scenario to the homogeneous and isotropic
FRW brane leads to modification of
 the Friedmann equations with a quadratic correction to
 energy density at higher energies.  Many issues in cosmology like inflation, dark
 energy, cosmological constant were investigated in the braneworld cosmological scenario  and interesting results
 were obtained.  Later the studies were extended to
  anisotropic brane, such as cosmological solutions to the Bianchi-I
 and V braneworld models were obtained and it is shown that for matter obeying barotropic equation of state, there could be an intermediate phase
  of high anisotropy which in turn leads to an isotropic phase in the later
  universe \cite{campo}.  Another interesting result is obtained in \cite{varun},  with the scalar field as source, a large initial
anisotropy induces significant damping in the dynamics of the scalar field   resulting in
greater inflation. The shear dynamics on Bianchi I brane cosmological model with a perfect fluid is considered and shown that shear parameter has a maximum during a transition period from non-standard to standard cosmology \cite{top}.

Motivated by these developments, we consider
evolution of the density parameter in an anisotropic DGP braneworld model and also examine the late time acceleration in this model.   Anisotropic brane in the DGP model is also
considered in \cite{riz,sur} and solutions are obtained without affecting the basic features of the model.
The DGP cosmological model possesses two classes of solutions; one which is close to the standard FRW cosmology and the other  is either a fully five dimensional regime or a self-inflationary solution which produces accelerated expansion. The self-accelerated branch is known to contain ghost and tachyonic like excitations \cite{ant,padila,charm,gorb,koy}, but nevertheless its ability to explain the late time acceleration makes it an interesting case for study and has generated enormous amount of interest in the recent years. We mainly focus on the
effects of shear and cross over scale on the dynamics of
dimensionless density parameter and the deceleration parameter.
 The evolution equation for $\Omega_\rho$ is derived in the self-accelerated branch and  for the
fully five dimensional branch  and the corresponding deceleration parameter are also obtained.

\section{DGP dynamics on Bianchi-I brane}
In the DGP model our universe is a 3-brane embedded in
five-dimensional (5D) Minkowski bulk and there is an induced four-dimensional
Ricci scalar on the brane, due to radiative correction to the
graviton propagator on the brane \cite{dvali}. In this model
gravity is modified at large length scales and there is a crossover
scale between the four-dimensional and five-dimensional
gravity given by  $$r_c =\frac{k^2_5}{2\mu^2}$$
 where $k^2_5$ and $\mu^2$ are the 4D and 5D gravitational constants respectively.
 Below the crossover scale $r_c$, the
potential has the usual Newtonian form and above which the gravity
becomes five-dimensional (A similar scenario with a scalar curvature term in the brane action is considered in \cite{holdom,colins} with a brane embeded in AdS bulk. Here, it is shown that it is possible to generate a four-dimensional theory
of gravity depending upon AdS lengths). We start with a generalized DGP model in
which both bulk cosmological constant $\Lambda$ and brane tension
$\sigma$ are non-zero. The effective Einstein equation on the
brane can be written  \cite{riz} as
\begin{eqnarray}\label{gendgp}
 \left( 1+\frac{\sigma k^2_5}{6 \mu^2}\right )
G_{\mu\nu}&=&-\left(\frac{k^2_5
\Lambda}{2}+\frac{k^4_5\sigma^2}{12}\right
)q_{\mu\nu}+\mu^2 T_{\mu\nu} \nonumber  \\
 &&     +\frac{\sigma k^4_5}{6}\tau_{\mu\nu} +\frac{k^4_5}{\mu^4}
F_{\mu\nu} + k^4_5 \Pi_{\mu\nu} + \frac{ k^4_5}{\mu^2}
L_{\mu\nu}-E_{\mu\nu},
\end{eqnarray}

\begin{equation}
\Pi_{\mu\nu} = -\frac{1}{4}\tau_{\mu\rho}\tau^\rho_\nu
+\frac{1}{12}\tau \tau_{\mu\nu}+\frac{1}{8} q_{\mu\nu}
\tau_{\alpha\beta}\tau^{\alpha\beta}-\frac{1}{24}q_{\mu\nu}
\tau^2,
\end{equation}

\begin{equation}
F_{\mu\nu} = -\frac{1}{4}G_{\mu\rho}G^\rho_\nu+\frac{1}{12} G
G_{\mu\nu}+\frac{1}{8} q_{\mu\nu}
G_{\alpha\beta}G^{\alpha\beta}-\frac{1}{24}q_{\mu\nu} G^2,
\end{equation}

\begin{equation}
L_{\mu\nu} = \frac{1}{4}(G_{\mu\rho}\tau^\rho_\nu + \tau_{\mu\rho}
G^\rho_\nu) - \frac{1}{12} (\tau G_{\mu\nu} + G \tau_{\mu\nu}
)-\frac{1}{4} q_{\mu\nu}
(G_{\alpha\beta}\tau^{\alpha\beta}-\frac{1}{3} G \tau),
\end{equation}
 where $\tau_{\mu\nu}$ is the energy-momentum tensor of the matter fields on the brane, $\sigma$ is the brane tension and $T_{\mu\nu}$ is the bulk energy-momentum tensor. It can be seen that apart  from the quadratic matter field corrections $\Pi_{\mu\nu}$ to energy-momentum, there are corrections coming from the induced curvature term through
$F_{\mu\nu}$ and $L_{\mu\nu}$. $E_{\mu\nu}$ are  the  non-local corrections, coming from extra dimensions which correspond to the projection
of the bulk Weyl tensor on the brane. If we define the four
velocity comoving with matter as $u_\mu$, the non-local term takes
the  following form, \be\label{weyl}
 E_{\mu\nu}= -4 r_c^2[U(u_\mu u_\nu +\frac{1}{3}h_{\mu
\nu})+P_{\mu\nu} +Q_\mu u_\nu + Q_\nu u_\mu].
 \ee
Where $h_{\mu\nu} = g_{\mu \nu} + u_\mu u_\nu $  and

\be U=-\frac{1}{4 r_c^2} E_{\mu\nu}u^\mu u^\nu, \ee is the
effective non-local energy density on the brane.

\be P_{\mu\nu}=-\frac{1}{4 r_c^2} \left[ h_\mu{}^\alpha
h_\nu{}^\beta-{\textstyle{1\over3}}h^{\alpha\beta}
h_{\mu\nu}\right]  E_{\alpha\beta} \ee is the effective non-local
anisotropic stress and,
 \be
  Q_\mu= \frac{1}{4 r_c^2} h_\mu{}^\alpha
E_{\alpha\beta}u^\beta\,, \ee represents the effective non-local
energy flux on the brane.
The brane energy momentum satisfies the conservation equations  $\nabla^\mu\tau_{\mu\nu}=0$, the detailed description of conservation and dynamical equations is given in \cite{roy}.

We are interested in anisotropic cosmological models on the brane, hence we consider the Bianchi I brane model. We start with the 5D metric of the following form
\begin{equation}
ds^2 =dy^{2}+q_{\mu\nu}dx^{\mu}dx^{\nu},
\end{equation}
where the brane is located at $y=0$. The Bianchi metric on the brane is
given by,
\begin{equation}\label{bianchi1}
ds^2|_{y=0} =q_{\mu\nu}(y=0)dx^{\mu}dx^{\nu}= -dt^2 + a_i^2(t) {dx^i}^2\,.
\end{equation}

The conservation equation  takes the following form,
\begin{equation}\label{cons}
\dot{\rho}+\Theta(\rho + p)=0
\end{equation}
\begin{equation}
D^\nu P_{\mu\nu} =0  \end{equation}
 \begin{equation}\label{cons2}
 \dot{U}+{\textstyle{4\over3}}\Theta{ U}+\sigma^{\mu\nu}{
P}_{\mu\nu}  =0
 \end{equation}

where dot denotes the $u^\nu\nabla_{\nu} $ and $\Theta$ represents
the volume expansion rate and $\sigma^{\mu\nu}$ is the shear
rate. There is no evolution equation for $P_{\mu\nu}$, since it is bulk degree of freedom and cannot be predicted from the brane \cite{roy}. The Hubble parameters for the Bianchi I metric is given by
$H_i = \frac{{\dot a_i}}{a_i}$ and one can define the mean
expansion factor as $S = (a_1a_2a_3)^{1/3}$, thus  \be \Theta
\equiv 3H = 3\frac{{\dot S}}{S} \equiv \sum_{i} H_i .\ee

 The modified Raychaudhuri equation for the Bianchi-I metric on the brane is obtained as \cite{riz},
\begin{eqnarray}
\dot{\Theta}+\frac{1}{3}\Theta^2 +\sigma^{\mu\nu}\sigma_{\mu\nu} &
=& \frac{-k^2_5}{2}\left[ \frac{1}{12 \mu^4}\left(3 \left( {
H^2+\frac{K}{S^2}} \right) -\mu^2\rho\right)^2 \nonumber
+\frac{1}{4}\rho(\rho+2p)   \right. \\  && \left.  - \frac{1}{2
\mu^2}\left( 3 \left( { H^2+\frac{K}{S^2}}\right) (\rho+p)
+G_{ii}\rho \right) \right. \nonumber  \\  && \left. +
 \frac{1}{4 \mu^4}3 \left( H^2+\frac{K}{S^2}\right) \left(3\left(  H^2+\frac{K}{S^2}\right)  +2G_{ii}\right) +\frac{8r_c^2}{k^4_5}U \right].
\end{eqnarray}

and Gauss-Codacci eqns are, \be \label{puv}
\dot{\sigma}_{\mu\nu}+\Theta\sigma_{\mu\nu}=4 r_c^2  P_{\mu\nu}\,
\ee

\be\label{cod} \frac{2}{3}\Theta^2 +\sigma^{\mu\nu}\sigma_{\mu\nu}
 = \frac{k^4_5}{6\mu^4}\left(3\left( H^2+\frac{K}{S^2}\right) -\mu^2\rho\right)^2  + 4 r_c^2 U\, .
\ee

The system of equations are not closed on the brane due to the presence of anisotropic stress term $P_{\mu\nu}$ and complete bulk solutions
are needed to determine brane dynamics. This problem is avoided(as in earlier papers  \cite{varun,riz,roy}) by choosing a special case $U=0$, this vanishing of non-local energy density leads
via (\ref{cons2}) to condition $\sigma_{\mu\nu}P^{\mu\nu}=0$. This consistency condition implies a condition on evolution of $P_{\mu\nu}$ as  $\sigma^{\mu\nu}\dot{{
P}_{\mu\nu}}= 4 r_c^2 P^{\mu\nu} P_{\mu\nu}$, following from Eq.
(\ref{puv}). As there is no evolution equation for $P_{\mu\nu}$ on
the brane, this is  consistent on the brane(for complete discussion see \cite{riz,roy}). This allows us to find the dynamics of shear scalar, equation (\ref{puv}) can be contracted with shear and integrated to get,
\begin{equation}\label{s}
\sigma^{\mu\nu}\sigma_{\mu\nu}  = 2\sigma^2 = {6\Sigma^2\over
S^6}\,,~~\dot\Sigma=0\,.
\end{equation}

The modified Friedmann type equation in the DGP cosmology for the
Bianchi-I  metric (\ref{bianchi1}) is obtained using eqns (\ref{cod}) and (\ref{s}) as
 \be\label{dgp}
 (H^2+\frac{K}{S^2}) -\frac{\Sigma^2}{S^6}-\frac{2 \mu^2}{k^2_5}\epsilon \sqrt{(H^2+\frac{K}{S^2})-\frac{\Sigma^2}{S^6}} =\frac{\mu^2}{3}\rho, \ee

where $\epsilon=\pm 1$ corresponds to two possible embedding of
the brane in the bulk space-time and gives rise to the two branches of
solutions in the late universe. We can rewrite
 (\ref{dgp}) as

\begin{equation}\label{dgpfred}
   H^2 +\frac{K}{S^2}=\left(\epsilon \frac{1}{2r_c} + \sqrt{\frac{1}{4 r^2_c} + \frac{\mu^2 \rho}{3}}  \right)
   ^{2} + \frac{\Sigma^2}{S^6},
\end{equation}
and define the following dimensionless variables
\begin{eqnarray}\label{omega}
  \Omega_{\rho} &=& \frac{\mu^2 \rho}{3 H^2}\\
  \Omega_{r_c} &=& \frac{1}{ 4H^2 r_{c}^2}\\
 \Omega_{K} &=& -\frac{K}{ H^2 S^2}\\
\Omega_{\Sigma} &=& \frac{\Sigma^2}{ H^2 S^6}.
\end{eqnarray}

 The Friedmann type equation (\ref{dgpfred}) takes the form

\begin{equation}\label{dgp norm}
1 =\left(\sqrt{\Omega_{r_c}} + \sqrt{ \Omega_{r_c} + \Omega_\rho }\right)^{2} + \Omega_{\Sigma}+ \Omega_{K}
\end{equation}
which implies the variables must take values in the interval [0,1]
and further it is seen  that the normalisation condition is
modified from the usual FRW case. For the $K=0$ isotropic DGP model, people
have estimated  values of $\Omega_{r_c}$ and $\Omega_{\rho}$ and  compared with observation and which are
 in good agreement with theory \cite{tuoma}.
For the Bianchi I case we have to take into account $\Omega_{\Sigma}$ to get
correct estimate of values of $\Omega_{r_c}$ and
$\Omega_{\rho}$, as can be seen from (\ref{dgp norm}), except in the limit $S\rightarrow \infty$ when effects of shear are negligible.

\begin{center}
\includegraphics[scale=0.65]{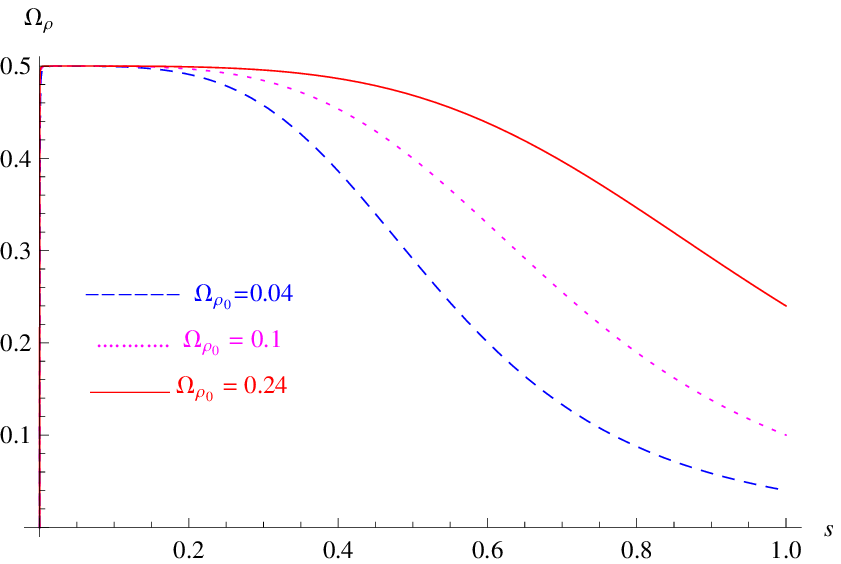}\\
{\small Figure 1.Evolution of $\Omega_{\rho}$ for  $\epsilon= +1$, for
different values of $\Omega_{\rho_0}$,  with $\Sigma=0.001$, $K=0, \gamma=4/3$ and $r_c=0.1$.}
\end{center}

\section{ Evolution of cosmological density parameter}
Next we  study the evolution of density parameter
$\Omega_{\rho}$ and see its dependence on shear $\sigma$ and cross over scale $r_c$. For
this we consider the two branches of solutions ($\epsilon=\pm 1$)
of the DGP scenario separately. The branch $\epsilon=+1$ corresponds to  the
self-accelerating solution in the late universe, without a
cosmological constant. Expanding equation (\ref{dgpfred}) under the
condition $ \mu^2 \rho r_c^2 \ll 1$ and considering $\epsilon=+1$
case up to first order \cite{riz,gumj}, we obtain the Friedmann
equation of the form,
\begin{equation}\label{late1}
H^2+\frac{K}{S^2}= \frac{1}{r_c^2}+ \frac{2}{3}\mu^2 \rho +
\frac{\Sigma^2}{S^6}.
\end{equation}
Using the definition (\ref{omega}) the above equation can be written as,
\begin{equation}\label{condi1}
H^2(2\Omega_{\rho}-1)= -\frac{1}{r_c^2} -
\frac{\Sigma^2}{S^6}+\frac{K}{S^2},
\end{equation}
it can be noted from the above equation that $K=0$ no longer implies that $\Omega_{\rho}=1$, as in the standard FRW case, we see the extra dependence on $r_c$ which is a typical feature of the DGP model. Also the effects of shear $\Sigma$ cannot be neglected and
value of $\Omega_{\rho}$ is constrained by these parameters. We also derive the following quantity and  will  use it later
\begin{equation}\label{diff fred}
\dot{H} = -H^2-\mu^2(\rho +p) + \frac{1}{r_c^2} +
\frac{2}{3}\mu^2 \rho -2 \frac{\Sigma^2}{S^6}.
\end{equation}

\begin{center}
\includegraphics[scale=0.65]{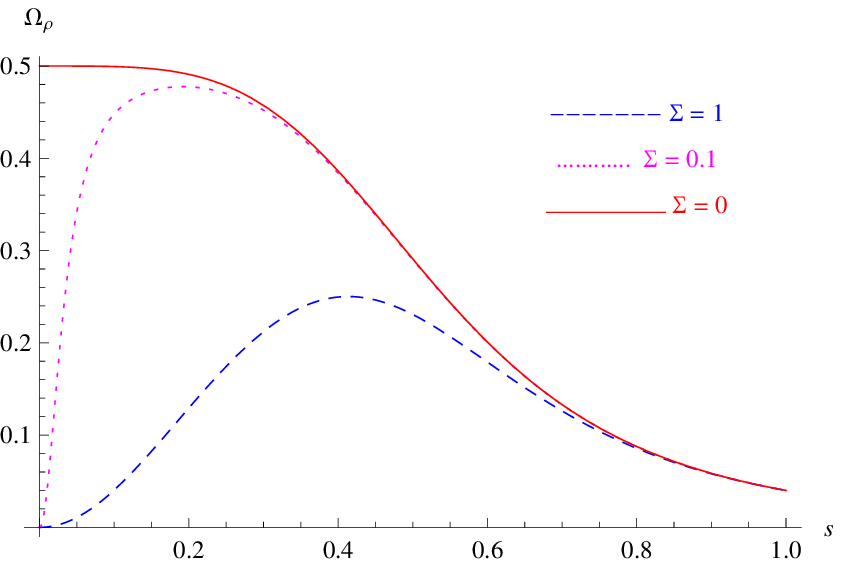}\\
\end{center}
{\small Figure 2. Evolution of $\Omega_{\rho}$ for $\epsilon= +1$ in the
isotropic case ( $\Sigma= 0$) and for different values of $\Sigma$, with $\Omega_{\rho_0}= 0.04$, $K=0$,
$\gamma=4/3$ and $r_c=0.1$.}

In order to study the evolution of $\Omega_{\rho}$, we differentiate eqn
(\ref{omega}) and use eqns (\ref{late1}) and (\ref{diff fred}) to   get,
\begin{equation}\label{om2}
    \frac{d\Omega_\rho}{d\tau}=  H \Omega_\rho \left[ (3\gamma-2)(2\Omega_\rho-1)-\frac{1}{H^2}\frac{2}{r_c^2}+
\frac{1}{H^2} \frac{4 \Sigma^2}{S^6}\right].
\end{equation}

Where we have assumed that the total energy density and pressure
are related by linear barotropic equation of state,
$p=(\gamma-1)\rho $. Next, to obtain equation for the $\Omega_\rho(S)$ phase plane analysis  we divide  eqn (\ref{om2}) by
$\frac{dS}{d\tau}=HS$ and substituting $H$ from (\ref{condi1}) we
get,

\begin{equation}\label{om3}
\frac{d\Omega_\rho}{dS}= \frac{\Omega_\rho}{S}
\left[(3\gamma-2)(2\Omega_\rho-1) +(2\Omega_\rho-1)
\frac{-\frac{2}{r_c^2}+ \frac{4 \Sigma^2}{S^6}}{-\frac{1}{r_c^2}-
\frac{\Sigma^2}{S^6}+\frac{K}{S^2}} \right].
\end{equation}

The above equation can be integrated to get,

\begin{equation}\label{om1}
\Omega_\rho=\left(1+\frac{1-2
\Omega_{\rho_0}}{\Omega_{\rho_0}}\frac{S^6-3KS^4r_c^2+\Sigma^2r_c^2}{S_0^6-3KS_0^4r_c^2+\Sigma^2r_c^2}\left(\frac{S}{S_0}\right)^{3\gamma-6}\right)^{-1},
\end{equation}
here $\Omega_{\rho_0}$ and $S_0$ denote the values of
density parameter and scale factor at the present time.
It is seen from (\ref{om1})  that evolution equation depends on the shear and its behaviour is significantly effected in the early universe, depending on the value of $\Sigma$. And the value of $\Omega_{\rho}$ as  $S\rightarrow 0$ is around 0.5 for all values of $\gamma$ in the range (0,2), whereas in the FRW it depends on the $\gamma$ and equation of state. We analyse this further by giving phase plane plots of $\Omega_{\rho}$,
Figure.1 shows  behaviour of $\Omega_{\rho}$ as a function of $S$ for different
values of $\Omega_{\rho_0}$ in  the self accelerated branch
of the solution. The value of  $\Omega_{\rho}$ for $S$ very close to zero is same for different values of $\Omega_{\rho_0}$ and starts deviating  for large values of $S$.
Figure 2 shows the  comparison  of $\Omega_{\rho}$ in the
isotropic case \footnote{For plotting the isotropic case we have used the expression $\Omega_\rho=\frac{1}{1+\frac{1-2
\Omega_{\rho_0}}{\Omega_{\rho_0}}\frac{S^2-K^2r_c^2}{S_0^2-K^2r_c^2}\left(\frac{S}{S_0}\right)^{3\gamma-6}}$.} along with different values of $\Sigma$ in the anisotropic case. It is observed that for the isotropic case  $\Omega_\rho$ starts with a high value, whereas for anisotropic case i.e. $\Sigma=0.1,1$ the  plot is peaked at intermediate value of $S$ and starts decreasing for large values of $S$.  The dependence of density parameter on $\gamma$ is shown in the figure 3, the asymptotic value of $\Omega_\rho$ as $S\rightarrow 0$ is around 0.5 irrespective of the value of $\gamma$. And the density parmater decays faster for small values of $\gamma$. Finally the effects of cross over scale on the evolution of cosmological density parameter is shown in
 Figure 4. We have shown that changing the value of crossover scale i.e. $r_c=0.1,1,2$ does not vary $\Omega_\rho$ much, whereas if we simultaneously change the value shear parameter $\Sigma$ to a higher value, the behaviour is significantly modified for the same values of $r_c$.

\begin{center}
\includegraphics[scale=0.65]{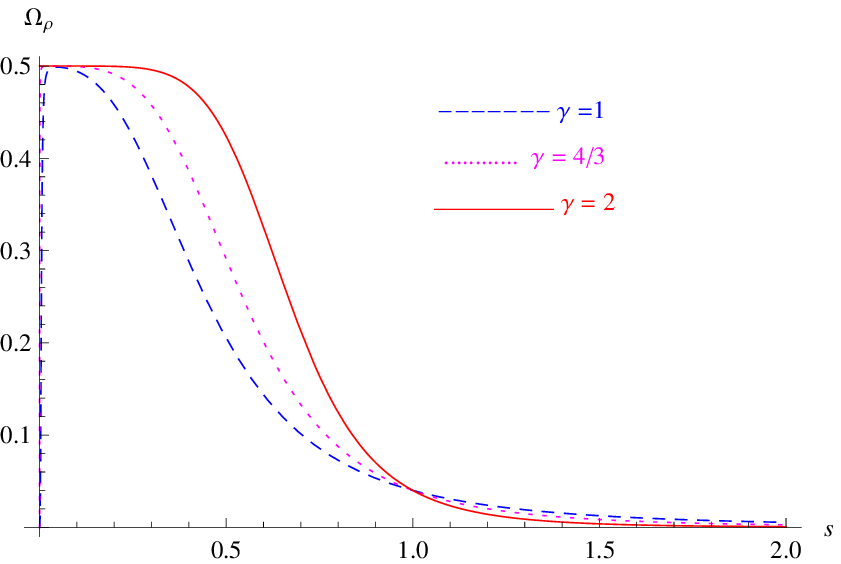} \\
\end{center}
{\small Figure 3. Evolution
of $\Omega_{\rho}$ for $\epsilon= +1$ with different values of
$\gamma$ and
$\Omega_{\rho_0}= 0.04$, $K=0$,  $\Sigma= 0.001$, $r_c=0.1$.}

We can define the following dimensionless deceleration parameter  $q=-
\frac{\ddot{S}S}{\dot{S}^2}$ or $q=\frac{\dot{H}+ H^2}{H^2}$  to study the accelerated phase of the universe, which can be written using
(\ref{diff fred}) as,
\begin{equation}\label{q1}
    q=(3\gamma-2)\Omega_\rho - (2\Omega_\rho-1)\frac{S^6-2 \Sigma^2 r_c^2}{-S^6- \Sigma^2 r_c^2 + K S^4}\,.
\end{equation}
Acceleration takes place when $q<0$, which is determined in the standard FRW case  by the value of $\gamma$ and $\gamma=2/3$ forms the critical value. It is clear from equation (\ref{q1}) in this case entry into the accelerating phase is determined by values of the shear and  cross-over scale. Figures 5 shows the behaviour of deceleration
parameter as a function of scale factor, Fig.5(a) shows the
dependance of $q$ on the  shear and it is seen that higher value of the
shear  imply that the universe enters an accelerated phase at later time. In figure 5(b) the decceleration parameter is plotted for different values of $r_c$, it is seen the irrespective of the value of cross over scale, the universe enters a self accelerated phase at the same time. And in both the cases (5(a) and 5(b)) it is noted that in the limit $S\rightarrow 0$,  $q$ approches a value 2.

\begin{center}
(a) \includegraphics[scale=0.65]{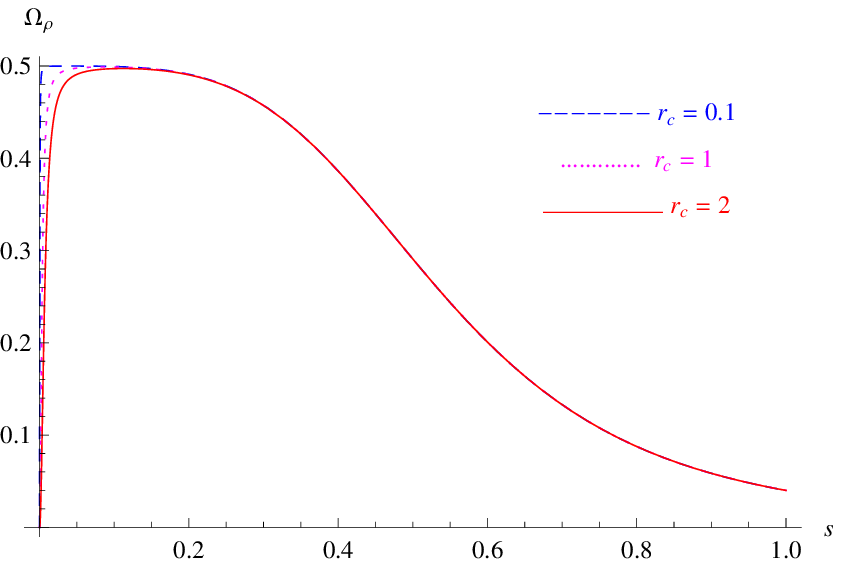}
(b) \includegraphics[scale=0.65]{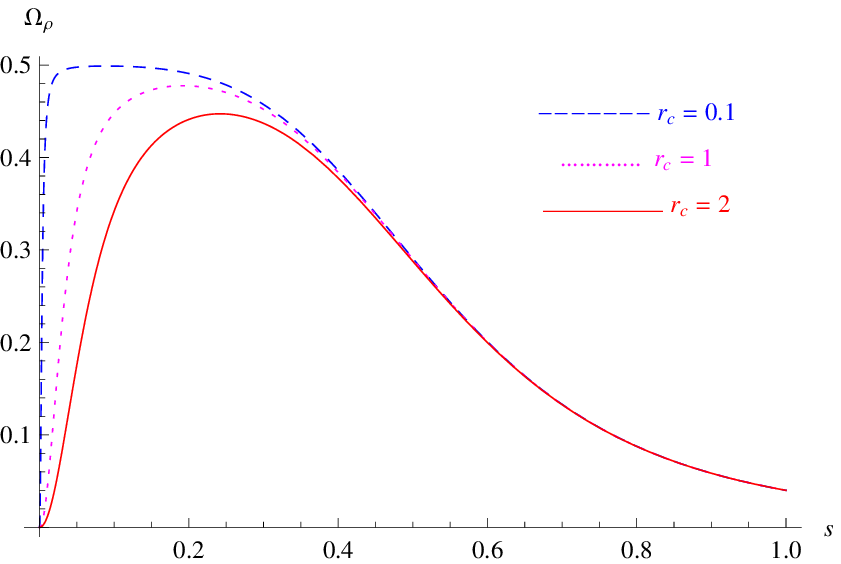}\\
\end{center}
{\small Figure 4. Evolution
of $\Omega_{\rho}$ for $\epsilon= +1$ for different values of
$r_c$ and $\gamma=4/3$,
$\Omega_{\rho_0}= 0.04$, $K=0$. In the left panel we have taken the value of $\Sigma=0.001$ and in the right panel $\Sigma=0.01$, to show that a higher value of shear will lead to large variation in $\Omega_{\rho}$ with $r_c$. }

Next, we consider the $\epsilon=-1$ case (which is also known as the
fully 5D case \cite{def}), again expanding (\ref{dgpfred}) with
condition $ \mu^2 \rho r_c^2 \ll 1$ we get  the Friedmann type
equation as,
\begin{equation}\label{fred2}
 H^2 +\frac{K}{S^2} =  \frac{\mu^4 r_c^2}{9}
\rho^2 + \frac{\Sigma^2}{S^6}.
\end{equation}
Since it represents the fully 5D case, the Freidmann equation has only quadratic contribution from the energy density. We define new
dimensionless parameter as
 \begin{equation}\label{omega1}
\Omega_{\rho^2}=\frac{\mu^4 \rho^2 r_c^2}{9 H^2},
\end{equation}
this allows us to rewrite the equation (\ref{fred2}) in the form
\begin{equation}\label{late2}
H^2(\Omega_{\rho^2}-1)= \frac{K}{S^2}-\frac{\Sigma^2}{S^6},
\end{equation}
and
\begin{equation}\label{accl2}
\dot{H}=- H^2 -\frac{\mu^4 r_c^2\rho}{9} (2\rho +3p)-2
\frac{\Sigma^2}{S^6}.
\end{equation}

Next, we discuss  evolution equation for the dimensionless density parameter, for this we
follow the method used in $\epsilon=+1$ case. Differentiating
(\ref{omega1}) with respect to time and using equations
(\ref{late2}) and (\ref{accl2}) we obtain,
\begin{equation}\label{om4}
 \frac{d\Omega_{\rho^2}}{d\tau}=  H (\Omega_{\rho^2}-1)\Omega_{\rho^2} \left[2 (3\gamma-1)+ \frac{4 \Sigma^2 }{S^4 K-\Sigma^2}\right].
\end{equation}

Integrating  equation (\ref{om4}) gives,
\begin{equation}
\Omega_{\rho^2}=\left(1+\frac{1-
\Omega_{\rho^2_0}}{\Omega_{\rho^2_0}}\left[\frac{KS^4-\Sigma^2}{KS_0^4-\Sigma^2}\right]\left(\frac{S}{S_0}\right)^{6(\gamma-1)}\right)^{-1},
\end{equation}
where $\Omega_{\rho^2_0}$ is the value at the present time. Let us analyse the effects of shear on the behaviour of $\Omega_{\rho^2}$. Fig.6(a) shows evolution of $\Omega_{\rho^2}$
for the the $\epsilon=-1$ case, it is seen that asymptotic value of $\Omega_{\rho^2}$ for $S\rightarrow 0$ is around 1 for all values of $\Omega_{\rho^2_0}$. But, if we take different values of $\gamma$ the behaviour changes drastically, figure 6(b) shows $\Omega_{\rho^2}$ for different values of $\gamma$ for $K=0$. It is observed that for $\gamma=4/3,2$ in the limit $S\rightarrow 0$ the value of $\Omega_{\rho^2}$ is around 1 and decays from there. On the other hand for $\gamma=2/3$  in the limit $S\rightarrow 0$ the value of $\Omega_{\rho^2}\rightarrow 0$  and increases for large values of $S$.

 Now the deceleration parameter can be obtained in
this case using (\ref{late2}) and (\ref{accl2}) as,
\begin{equation}\label{q2}
q=(3\gamma-1)\Omega_{\rho^2}  + 2(\Omega_{\rho^2}-1)\frac{\Sigma^2}{S^4K-\Sigma^2}.
\end{equation}

Here it is noticed that when $K=0$, entry in to the accelerated branch is determined by the value of $\gamma$. And, $\gamma=1/3$ forms the critical value that separates the accelerating phase from the deccelerating, compared to FRW case where $\gamma=2/3$ is the critical value.

\begin{center}
(a) \includegraphics[scale=0.65]{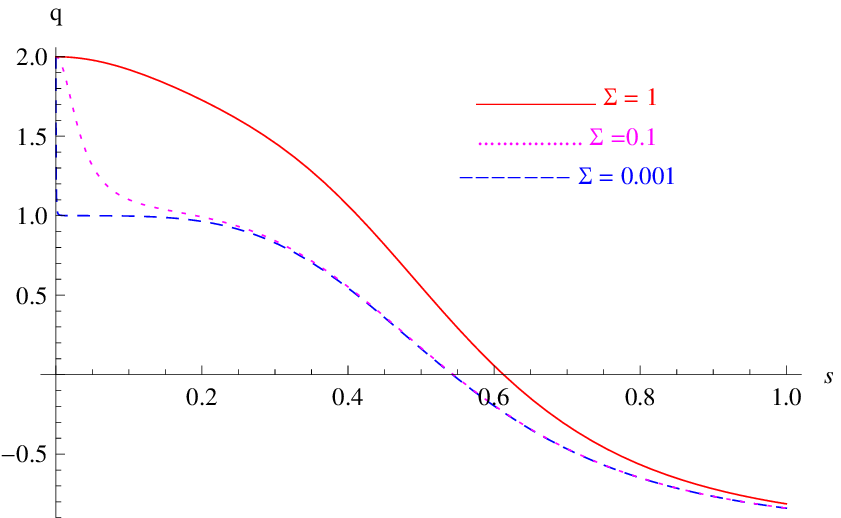}
(b)\includegraphics[scale=0.65]{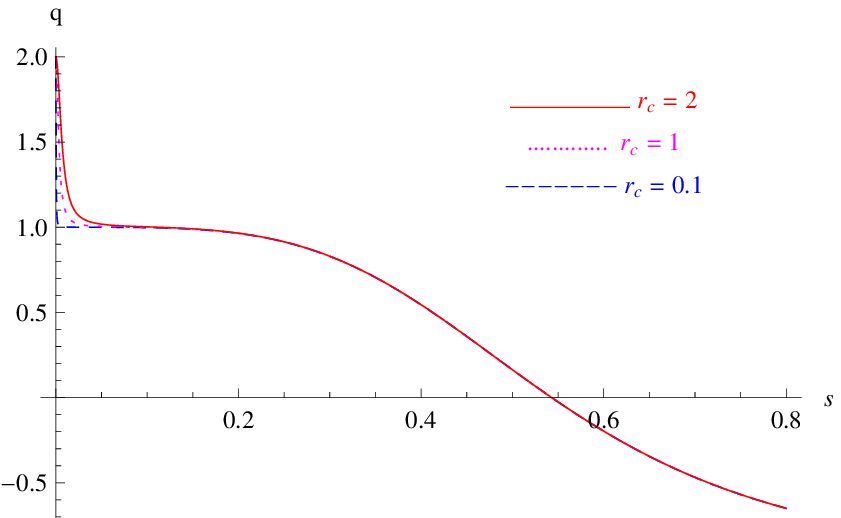}\\
\end{center}
{\small Figure 5. Plot for decceleration parameter $q$
for $\epsilon=+1$ (a) shows different values of shear parameter $\Sigma$ and $r_c=0.1$, (b) shows
evolution with different values of
$r_c$  and $\Sigma= 0.001$  with
$\Omega_{\rho_0}= 0.04$ ,$\gamma=4/3$, $K=0$ .}

\begin{center}
(a)\includegraphics[scale=0.65]{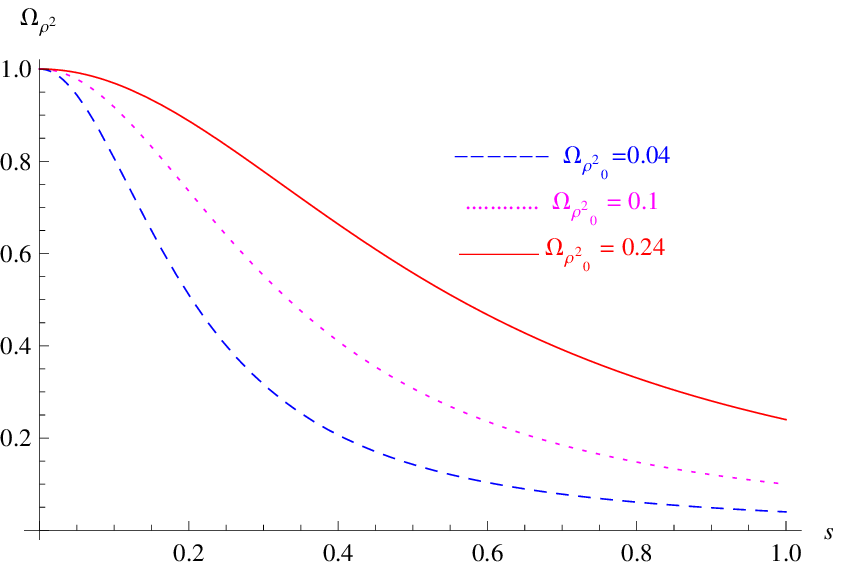} 
(b) \includegraphics[scale=0.65]{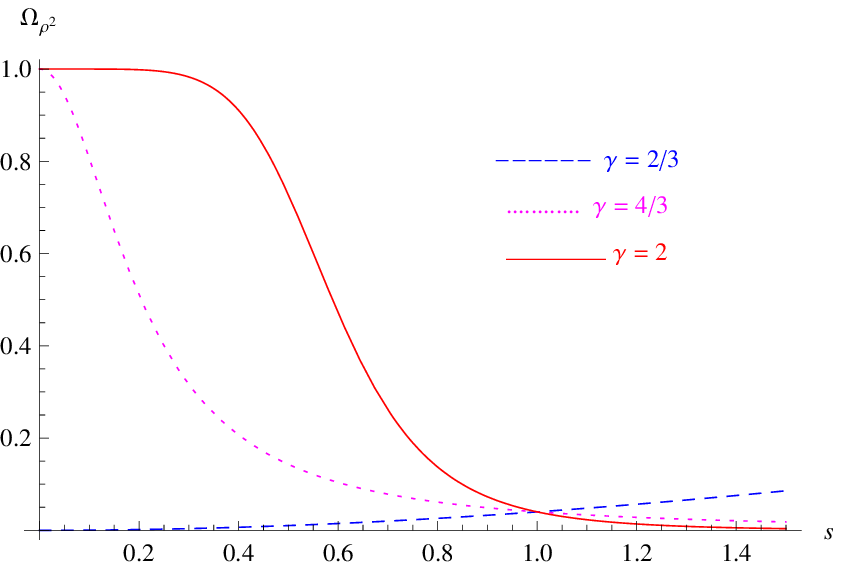}  \\
\end{center}
{ \small Figure 6. Figure (a) shows behaviour of $\Omega_{\rho^2}$ in $\epsilon=-1$ case for
different values of $\Omega_{\rho_0}$ and $\gamma=4/3$. (b) for different values of $\gamma$ and $\Omega_{\rho_0}=0.04$. In both the plots we use $\Sigma= 0.001$, $K=0$ and $r_c=0.1$.}

\section{Conclusions}
We  studied the evolution of cosmological density parameter in an anisotropic DGP braneworld model. The two branches
of solutions are considered separately but  we emphasized  on
 the self accelerated branch with shear parameter  and
crossover scale.
It is found that the evolution of dynamical density parameter is not altered much during the late  universe by the presence of shear. It is seen in figure
1 that asymptotic value of $\Omega_{\rho}$ as $S\rightarrow 0$ approaches 0.5  and decreases  for large values of $S$ . The behaviour of density parameter  for different values of shear is shown in figure 2 and is compared with the isotropic case. In the early universe shear plays a  significant role  and in the late universe evolution coincides with that of a isotropic DGP case. It is also observed that in the presence of shear irrespective of the value $\gamma$, the behaviour of $\Omega_{\rho}$ is same near
$S=0$. The density parameter $\Omega_{\rho}$ does not seem
to vary much with value of cross over scale, for a smaller value of shear, as shown in the figure 4(a). Increasing the value of shear to a higher value ($\Sigma=0.01$) the density parameter varies significantly, depending on cross over
scale, in the early universe. But in the late universe is insensitive to the value of $r_c$.  Deceleration parameter is plotted in figure
5, it shows that in the presence of shear universe enters  a self accelerating phase at later time, for high values of shear. It is observed that value of $r_c$ does not change the evolution of $q$ significantly and universe enters a self-accelerated phase as shown in figure 5(b).
Now, in the $\epsilon=-1$ or fully five dimensional branch of the solution, the density parameter shows interesting behaviour depending on the value of $\gamma$, in the limit $S\rightarrow 0$ the evolution is drastically changed for $\gamma=2/3$. Looking at the decceleration parameter in this case, we found that  $\gamma=1/3$ forms the critical value that divides the accelerating  phase from the deccelerating.  In this work we were interested in the $K=0$ case for both the branches of DGP model, the behaviour of $\Omega_{\rho}$ and $q$  would change significantly for the universe with a non-zero spatial curvature.

As the anisotropic effects are relevant at high redshifts and decreases with the expansion of the universe, we would like to estimate  the value of the shear at the last scattering surface.  We obtain the value of shear from our equations to be of the order   $\frac{\Sigma}{H_0} \sim 1.68 \times 10^{-10}$, this is in agreement with the limits obtained for the other Bianchi models. Hence, the estimated value of shear from our model is compatible with the late time evolution of DGP model and the crossover scale in this case is consistent with the one obtained by others\cite{Lomb,sohrab}.  Finally, we would like to conclude that the dynamical evolution of density parameter does not affect  late time behaviour of the DGP model in the anisotropic case and the results are in agreement with isotropic case in the late universe.

\end{document}